\documentclass{iopart}

\usepackage{iopams}
\usepackage{graphicx}

\begin{document}

\article[Fluid Dynamics as Diagnostic Tool]{Quark-Matter 2006}{Fluid
Dynamics as Diagnostic Tool for Heavy Ion Collisions}

\author{
Bj\"orn B\"auchle$^{1,2}$,
Yun Cheng$^1$,
L\'aszl\'o P Csernai$^{1,3}\footnote{Speaker}$,
Volodymyr K Magas$^{4}$,
Daniel D Strottman$^{5}$,
P\'eter V\'an$^{1,3}$ and 
Mikl\'os Z\'et\'enyi$^1$
}
\address{$^1$ Section for Theoretical Physics, Departement of Physics,
University of Bergen, All\'egaten 55, 5007 Bergen, Norway}
\address{$^2$ Institut f\"ur theoretische Physik, Universit\"at Frankfurt,
Max-von-Laue-Stra\ss e 1, D-60438 Frankfurt am Main, Germany}
\address{$^3$ KFKI Research Institute for Particle and Nuclear Physics, P.O.
Box 49, 1525 Budapest, Hungary}
\address{$^4$ Departament d'Estructura i Constituents de la Materia,
University of Barcelona, Av.\ Diagonal 647, 08028 Barcelona, Spain}
\address{$^5$ Theory Division, Los Alamos National Laboratory, Los Alamos,
NM, 87454, USA}

\eads{\mailto{csernai@ift.uib.no},
\mailto{baeuchle@th.physik.uni-frankfurt.de}}

\begin{abstract}

Ultra-Relativistic Heavy Ion Collisions at an energy $\sqrt{s_{NN}} = 65~{\rm
GeV}$ are studied in a three-dimensional Fluid Dynamical model. The results of
a hydrodynamical evolution using the PIC-method are shown. The importance and
diagnostic value of a proper reaction plane determination is emphasized, and
the time development of collective observables is presented.

\end{abstract}

 \submitto{\jpg}

\paragraph{Introduction:} Fluid Dynamical (FD) models are widely used to
describe ultra-relativistic heavy ion collisions. Their advantage is that one
can vary flexibly the Equation of State (EoS) of the matter and test its
consequences on the reaction dynamics and the outcome. In energetic collisions
of large heavy ions, especially if QGP is formed in the collision, one-fluid
dynamics is a valid and good description for the intermediate stages of the
reaction.  Here, interactions are strong and frequent, so that other models
have limited validity. On the other hand, the initial and final, Freeze-Out
(FO), stages of the reaction are outside the domain of applicability of the
fluid dynamical model.

The equations of perfect FD are just the local energy, momentum and baryon
number conservation laws, where we assume local equilibrium, so that the
energy-momentum tensor is given by the EoS. Thus, the most important and
almost only input is the EoS.

Of course the initial and final (FO-) conditions are also vital, and the
deviations from thermal equilibrium (viscosity and dissipation) are also
important. Here, though, we will concentrate on the impact of the initial
condition.

As pointed out above, the equations of perfect hydrodynamics are the
conservation laws. Still, the approach additionally assumes a local thermal
equilibrium. If this is present, intensive macroscopic space-dependent
quantities such as temperature, $T$, chemical potential, $\mu_B$, and
pressure, $P$, can be defined. An Equation of State has to be introduced,
which gives a relationship between energy- and baryon number density and the
pressure: $P = P \left ( e, \, n \right)$. This is necessary to have a
complete, solvable set of equations.  After introducing the shear viscosity,
$\eta$, and the heat conductivity, $\kappa$, a state near equilibrium can also
be described.

\paragraph{The PIC Model:} The Particle in Cell (PIC) method was developed by
Amsden and Harlow in the early 1960s \cite{AmsdenHarlow} and upgraded to
ultra-relativistic energies by Nix and Strottman in the 80s and 90s
\cite{Clare:1986qj}. The method takes advantage of a combination of Eulerian
and Lagrangian solution methods; the pressure is calculated on an Eulerian
grid, while all current transfers are calculated in a larger number of
Lagrangian cells, the so-called ``marker particles''. 

The method has numerous advantages (e.g.\ keeping conserved quantities
constant), and a typical main disadvantage, the development of ``ringing
instabilities'' in rapid expansion or explosion. This has recently been
eliminated by assigning the marker particles random positions to avoid
structures, along which an instability would start to develop. With this
upgrade the stability of expansion is increased, so that a spherical object
remains exactly spherical (within the numerical resolution) even after a
volume increase by a factor of more than 100.

After upgrades, the code can follow the fluid dynamical expansion even well
beyond the physical applicability of FD. This extended FD history can then be
used to analyse the effects of choices on the FO hyper-surface, or on the FO
layer (a 4-volume domain).

A code with finite cell-size exhibits necessarily a numerical viscosity and
coarse graining. The kinetic energy lost in this averaging is thermalized just
as in nature. No additional constraint is enforced to conserve the entropy.

The numerical viscosities are in the range of the estimated physical viscosity
\cite{Csernai:2005ny}, if the reaction is not too peripheral, in which case
the applicability of FD is not the best anyway. The larger than realistic
viscosities in the transverse direction dissipate more transverse flow to
heat, so we get higher $T$ and higher thermal smearing.

\paragraph{The initial State:}

\begin{figure}\begin{center} 
\includegraphics[width=6cm]{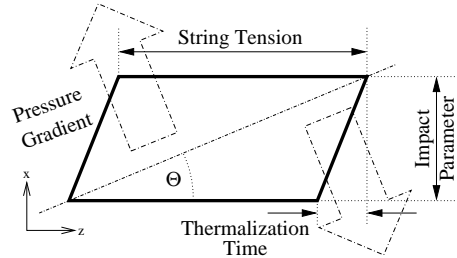}

\caption{A schematic view on the initial state of a heavy ion collision as
used in PIC. The labeled lengths indicate the dependence on the particular
quantity, not the quantity itself.} \label{fig:initialstate}

\end{center}\end{figure}

The initial shape is calculated in an effective 1D Yang-Mills coherent field
model, where the transverse plane is split up to ``streaks'' like in the
``Firestreak''-model of the 1970s \cite{Magas:2000jx,Magas:2002ge}.
Projectile and target matter interpenetrate each other initially in the
collision, exchange color charge and form ``strings'' which expand until all
initial kinetic energy is converted to string tension.  No transverse momentum
exchange is taken into account among the streaks in this initial stage, but a
longitudinal expansion at the edges of the strings was included in the
expansion to make the surface at the initial moment (when the FD starts)
smoother.

The streaks expand until the initial kinetic energy is converted fully into
string tension. We use an effective string tension of $6-8\,{\rm GeV/fm}$,
which was deduced from earlier string model fits to experimental CERN - SPS
results. This results in moderately long streaks of $4-8\, {\rm fm}$, so that
the thickness of the initial state is rather a flat than an extremely long
object parallel to the beam direction.

While in the middle of the transverse plane in the CM frame the matter is at
rest, the edges on the projectile and target side have projectile- and
target-directed excess momenta, which lead to a skewed object at the end of
the initial state. The longer the formation of initial state lasts, the more
skewed this initial state becomes (\fref{fig:initialstate}).

\paragraph{Non-existing Symmetry and Event-plane reconstruction:} Although
both nuclei involved in the collision of two gold beams are of equal size and
mass, there is no forward-backward (F/B)-symmetry. In each event a clear
distinction between projectile and target hemisphere must be made. It is
trivial to see that a single-event particle distribution following from an
initial state of finite impact parameter as described above, is not
F/B-symmetric but deviates from this by a certain angle $\Theta$ in the
x/z-plane (see \fref{fig:initialstate}). If the event-plane reconstruction is
not done for its F/B-orientation, these effects are lost.

The aforementioned effect, the \emph{third flow component}, has been predicted
as a signature of the QGP \cite{Csernai:1999nf,Csernai:2004gk}. It should be
visible in the special behaviour of the directed flow coefficient $v_1(y)$,
which is supposed to be closer to zero at central rapidities than the usual
linear increase of $v_1$ vs. rapidity, $y$. This effect is measurable at
around midrapidity \cite{Wang:2007kz}

Especially correlation measurements, but also flow determinations, must be
aware of this fact. Two-particle correlations showing basically two peaks at
$\Delta \eta = 0$ and $\Delta \phi = 0,\,2\pi$ do not take the real geometry
into account. Taking a sample over events with randomly identified F/B event
planes obscures the advent of interesting physics.

\paragraph{Results:}

\begin{figure}\begin{center} 

\includegraphics[width=.45\textwidth]{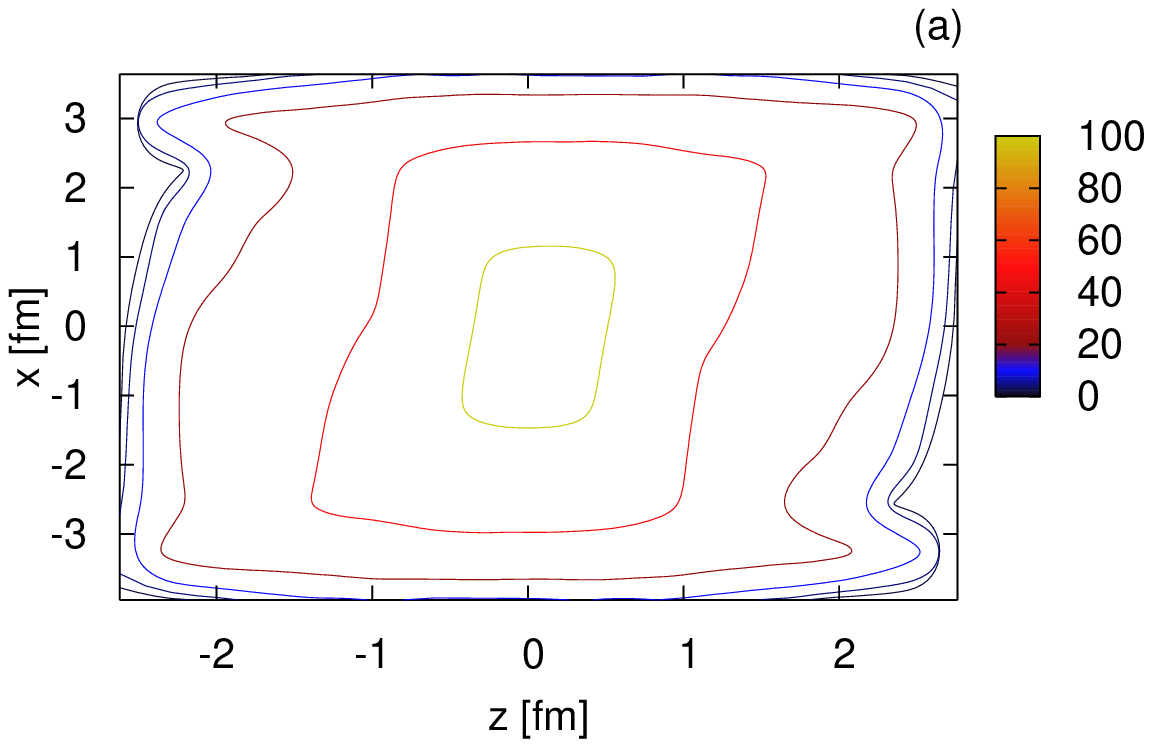}
\includegraphics[width=.45\textwidth]{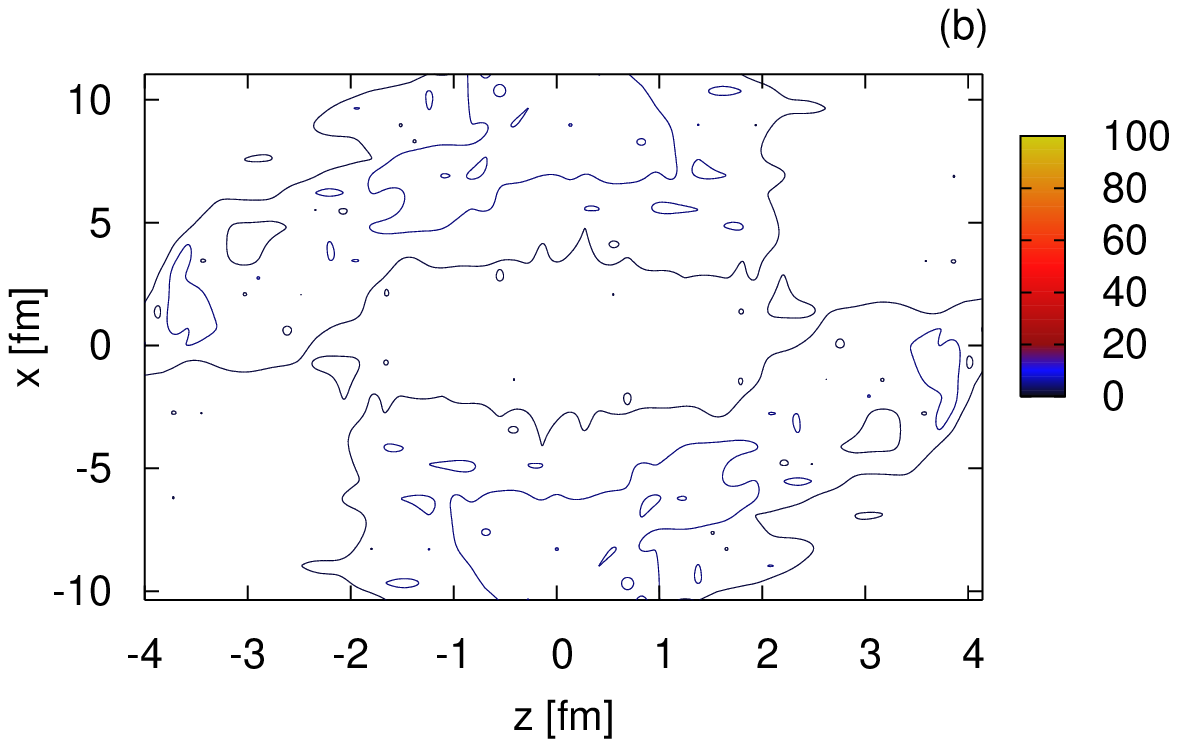} \caption{ (Color online)
The energy-density in the y = 0 -- plane at two different times (after
0.3~fm/c (a) and 2.7~fm/c (b)) in Au+Au-collisions at $\sqrt{s_{NN}} = 65~{\rm
GeV}$ at impact parameter $b = 0.7\,(R1 + R2 )$. Note different scales between
the plots.} \label{fig:energy}

\end{center}\end{figure}

\begin{figure}\begin{center} 

\includegraphics[width=.45\textwidth]{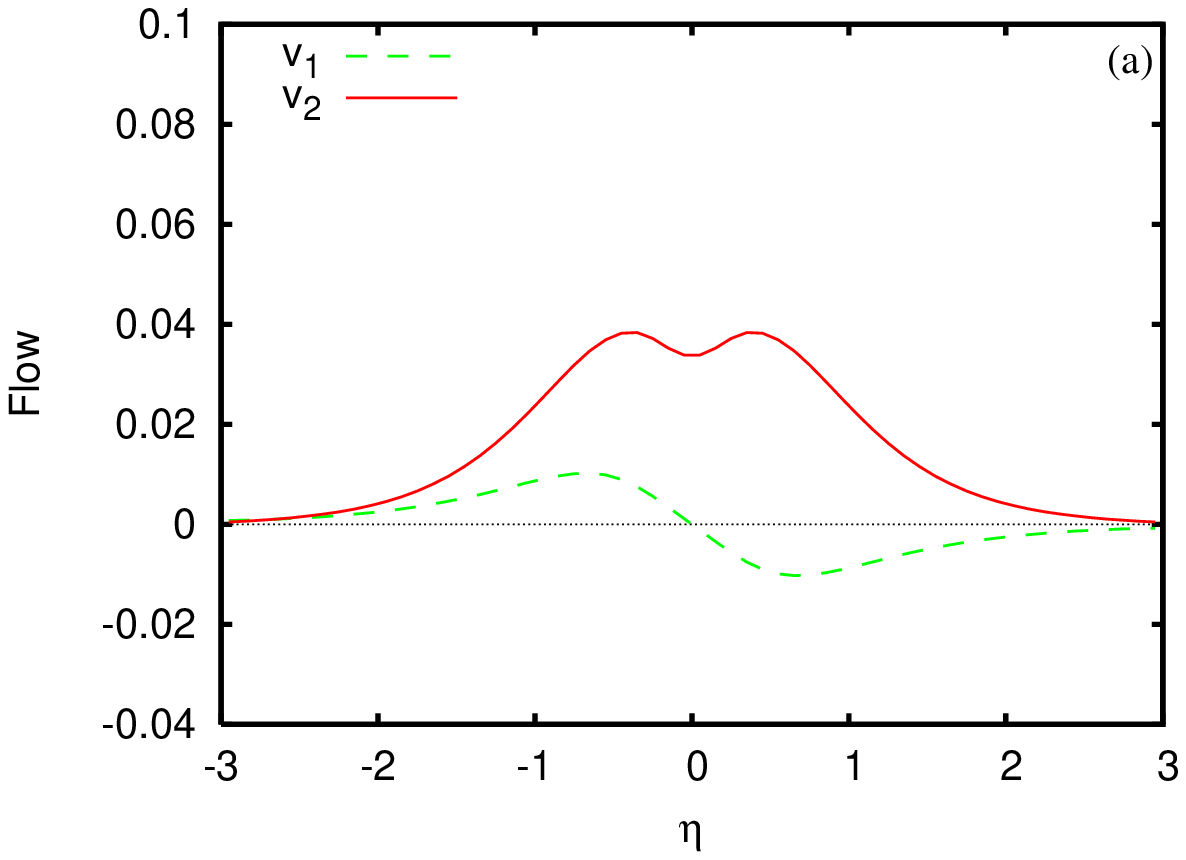}
\includegraphics[width=.45\textwidth]{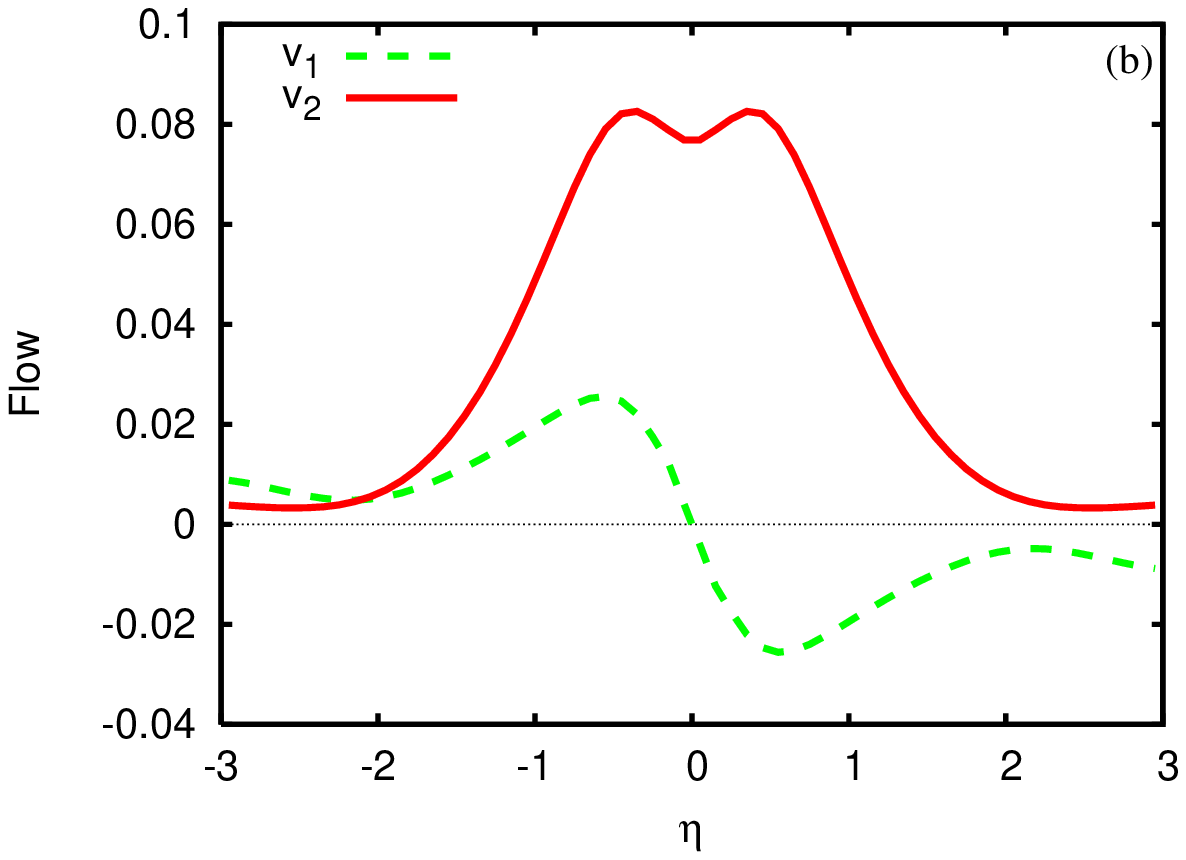}

\caption{(Color online) The flow coefficients $v_1$ and $v_2$ as a function of
pseudorapidity $\eta$ at two different times (after 0.3~fm/c (a) and 2.7~fm/c
(b)) in Au+Au-collisions at $\sqrt{s_{NN}} = 65~{\rm GeV}$ at impact parameter
$b = 0.7\,(R_1 + R_2)$.} \label{fig:flow:y}

\end{center}\end{figure}

The distorted initial state has the biggest pressure gradient along the lines
indicated in \fref{fig:initialstate}. This leads to a distribution of matter
and energy density in an ellipsoid shape in very late stages of the collision,
which is enhanced in the direction of the original pressure gradient.
\Fref{fig:energy} shows the calculated energy-density in the x/z-plane after 8
(a) and 71 (b) cycles, respectively (corresponding to times $t_1 = 0.3~{\rm
fm/c}$ and $t_2 = 2.7~{\rm fm/c}$ after the beginning of the hydro-evolution).
The flow patterns that follow from the initial state can be seen in
\fref{fig:flow:y}. The presence of the third flow component is well visible in
the former. It can also be seen that both directed and elliptic flow
coefficients approximately double their magnitude between the two times. 

All flow plots are calculated assuming massless quarks in a QGP-Phase. In
evaluating the flow during the hydro evolution we take advantage of the
observed $N_q$-scaling. It should further be noted that all flow plots are
calculated at constant time hypersurfaces. This is of course not including
freeze-out and hadronization, but it allows us to study effects of different
initial conditions during the hydro evolution. This leads to a larger $v_1$
than in experimental data \cite{Wang:2007kz}, because, we neither performed a
FO calculation through a realistic FO hypersurface, nor an impact parameter
averaging. Both these processes make the distributions smoother and decrease
$v_1$ and $v_2$.  The ``dip'' in $v_2$ at midrapidity is a consequence of too
small numerical viscosity in longitudinal direction.

\paragraph{Acknoledgements: } This work is supported by the Computational
Subatomic Physics project of the Research Council of Norway and by the Hadron
Physics EU Integrated Infrastructure Initiative.

\end{document}